\documentclass[journal]{IEEEtran}
\usepackage[utf8]{inputenc}

\title{End-to-End Demand Response Model Identification and Baseline Estimation with Deep Learning}
\date{}
\author{}

\usepackage{enumitem}
\usepackage{algorithmic}
\usepackage[libertine]{newtxmath}

\allowdisplaybreaks
\usepackage{float}
\usepackage{xifthen}
\usepackage{physics}
\usepackage{hyperref}
\usepackage{cleveref}
\usepackage{amsmath}
\usepackage{bbm}
\usepackage{amsthm,thmtools}
\usepackage{xcolor}
\usepackage{comment}
\usepackage{multirow}
\usepackage{graphicx}
\usepackage{bm}
\usepackage{cuted}
\usepackage[caption=false,font=footnotesize]{subfig}

\let\originalleft\left
\let\originalright\right
\renewcommand{\left}{\mathopen{}\mathclose\bgroup\originalleft}
\renewcommand{\right}{\aftergroup\egroup\originalright}

\newcommand{\bP}[2][]{\Pr\ifthenelse{\isempty{#1}}{}{_{#1}}\left[#2\right]}
\newcommand{\bE}[2][]{\mathop\mathbb{E}\ifthenelse{\isempty{#1}}{}{_{#1}}\left[#2\right]}
\newcommand{\bI}[2][]{\mathop\mathbb{I}\ifthenelse{\isempty{#1}}{}{_{#1}}\left[#2\right]}
\newcommand{\Var}[2][]{\mathbf{Var}\ifthenelse{\isempty{#1}}{}{_{#1}}\left[#2\right]}

\newcommand{\ud}[1]{_\mathrm{#1}}

\newcommand{\muo}[0]{\overline{\mu}}
\newcommand{\muu}[0]{\underline{\mu}}
\newcommand{\nuo}[0]{\overline{\nu}}
\newcommand{\nuu}[0]{\underline{\nu}}

\allowdisplaybreaks

\begin{document}
	
	\author{Yuanyuan~Shi,~\IEEEmembership{Member,~IEEE}, Bolun~Xu,~\IEEEmembership{Member,~IEEE}
		% \vspace{-.7cm}
		\thanks{Y.~Shi is with University of California San Diego, USA (email: yyshi@eng.ucsd.edu ); B.~Xu is with Columbia University, USA (email: bx2177@columbia.edu). }
	}  
	
	\maketitle
	\begin{abstract}
		This paper proposes a novel end-to-end deep learning framework that simultaneously identifies demand baselines and the incentive-based agent demand response model, from the net demand measurements and incentive signals. 
		This learning framework is modularized as two modules: 1) the decision making process of a demand response participant is represented as a differentiable optimization layer, which takes the incentive signal as input and predicts user's response; 2) the baseline demand forecast is represented as a standard neural network model, which takes relevant features and predicts user's baseline demand. 
		These two intermediate predictions are integrated, to form the net demand forecast.
		We then propose a gradient-descent approach that backpropagates the net demand forecast errors to update the weights of the agent model and the weights of baseline demand forecast, jointly.
		We demonstrate the effectiveness of our approach through computation experiments with synthetic demand response traces and a large-scale real world demand response dataset. Our results show that the approach accurately identifies the demand response model, even without any prior knowledge about the baseline demand.
		
		% The proposed method does not need prior knowledge about the baseline demand, and explicitly incorporates demand response disutilities and constraints into the learning model. \todo{A case study...}
	\end{abstract}
	
	
	\section{Introduction}
	
	Efficient utilization of demand-side resources in aiding grid operation is critical for a reliable and economic transition towards a low-carbon energy system. Federal Energy Regulatory Commission (FERC) has stated in the recent Order 2222~\cite{cartwright2020ferc} that wholesale electricity markets must ``level playing fields'' for behind-the-meter demand response (DR) resources, allowing them to be aggregated as a market participant under one or more market models. Future power system operators, utilities, and aggregators must be equipped with the new schemes and demand response models to fairly settle their market participation, and to predict their incentive-based behaviors associated with demand response instructions. 
	
	Yet, integration of DR resources comes with unique challenges~\cite{hu2015review}: the operator must be able to predict and measure how a DR participant responded, i.e., changes from its baseline demands, to incentive signals to ensure system reliability and fairly settle market payments. In practice, power system operators obtain direct output measurements from all participating resources and use these measurements to monitor the operation reliability and settle market payments for market participants. These on-the-meter grid resources react to grid instructions or price signals predictably based on their internal marginal production cost and physical operation constraints. However, these tasks become non-trivial over a DR participant: 1) a DR's response to market instruction is mingled with its baseline demand which cannot be metered in exact; 2) a DR may not have an explicit physical operation model or marginal production cost as the flexibility comes from an aggregation of appliances with human behavior involved. Although submetering could be used to improve the visibility over DR by installing meters inside a household to provide room or even appliance specific measurements, it poses serious privacy concerns, and the overwhelming amount of data could lead to a significant increase in the infrastructure investment which may out-weight the benefit of demand response~\cite{ji2016estimating, paterakis2017overview}.

	This paper proposes a novel DR parameter identification and baseline estimation approach using deep learning. We innovate the design of end-to-end model structure by having two modules for agent DR optimization and baseline demand forecast separately. For the agent DR optimization module, we incorporate differentiable optimization as a neural network layer motivated by Amos and Kolter's work~\cite{amos2017optnet}. Different from prior baseline estimation approaches using clustering~\cite{zhang2015cluster}, control groups~\cite{hatton2015statistical}, decomposition~\cite{hong2019exploration}, and self-report incentives~\cite{muthirayan2016mechanism},
	our method can identify the demand response model and the baseline demand directly from net demand measurements without any prior knowledge about historical baseline demands. Our work distinguishes itself from the existing literature in the following three aspects: 
	\begin{enumerate}
		\item We pioneer an end-to-end framework that jointly identifies a demand response agent model and its baseline demand. Our design leverages prior knowledge about the agent utility optimization model and reduces the computational complexity and the number of weights in magnitudes while providing good model explainability. We believe this could be a general recipe for end-to-end models to solve power system operation problems that involve both a prediction stage and a subsequent optimization stage.
		\item The proposed end-to-end learning framework is able to identify users' demand response, without any prior knowledge about the baseline demand. Compared to existing approaches that require historical baseline demand information, our approach obtains similar performance in agent model identification.
		\item We evaluate our model performance using synthetic and real-world demand response datasets showing the immediate applicability for real-world DR identification problems. In addition, we provide various ablation studies to demonstrate the robustness against measurement noises and time-varying agent models.
	\end{enumerate}
	The remaining of the paper is organized as follows: Section~II provides the background and literature review on demand response modeling. Section~III introduces the learning model and Section~IV explains the proposed training method. Section~V presents computation experiment results and Section~VI concludes the paper with a discussion on future directions.
	
	\section{Backgrounds and Literature}
	
	% \begin{figure}[t]%
	% 	\centering
	% 	\subfloat[]{
	% 		\includegraphics[trim = 0mm 00mm 0mm 0mm, clip, width = .95\columnwidth]{figures/dr_e1.eps}
	% 		\label{fig:sl1}%
	% 	}\\
	% 	\subfloat[]{
	% 		\includegraphics[trim = 0mm 00mm 0mm 0mm, clip, width = .95\columnwidth]{figures/dr_e2.eps}
	% 		\label{fig:sl2}%
	% 	}\\
	% 	\subfloat[]{
	% 		\includegraphics[trim = 0mm 0mm 0mm 0mm, clip, width = .95\columnwidth]{figures/dr_e3.eps}
	% 		\label{fig:sl3}%
	% 	}
	%   \caption{Illustration of (a) baseline demand; (b) time-of-use tariff; (c) demand response and actual net demand visible to the aggregator.}
	%     \label{fig:dr}
	% \end{figure}
	
	\subsection{Baseline demand estimation}
	Estimation of demand baselines is a fundamental requirement to systematically incorporate demand response into power system operation and market settlements: the system operator must disaggregate the effective change in the demand in response to grid needs from the demand baseline to monitor the demand resource, settle compensations based on market prices or tariffs~\cite{mathieu2011examining}. % Fig.\ref{fig:dr} shows an illustration of baseline demand and response, system operators can obtain the effective demand response components by trying to accurately forecast the baseline demand, and then subtract it from the actual net demand measurements. 
	
	% Yet, verifying a demand response participant's baseline had it not responded to instructions is non-trivial. The most intuitive way is to let the participant report its baseline after the dispatch period, but there is no guarantee of honesty as the participant can falsely report to gain more payment. For example, a building with a 30~kW true baseline demand can falsely report 40~kW of baseline to get paid for demand response without actually exercising the contract. To this end, systematic baseline estimation methods must be employed to ensure market efficiency. 
	
	Existing baseline estimation approaches can be grouped into top-down or bottom-up approaches. Top-down approaches assume demand response ``buyers'' (system operators or utilities) estimate baselines, such as use conventional demand forecast or use specialized price-responsive demand models~\cite{fernandez2021forecasting}, or ex-post baseline re-estimations based on realized system and weather conditions through methods such as clustering~\cite{zhang2015cluster} and control groups~\cite{hatton2015statistical}. Similar baseline estimation methods have been implemented in the industry. The Scheduled Load Reduction Program (SLRP) operated by Pacific Gas \& Electric Company (PG\&E) calculates the baseline by averaging the load demand of the selected time periods in the ten previous normal operating days~\cite{pge}. Bottom-up approaches, on the other hand, require demand response participants, the ``provider'' of demand response, to report baselines a-prior to the dispatch period, accompanied with supporting tariff mechanisms that encourage participants to truthfully report baselines base on the best of their knowledge~\cite{muthirayan2016mechanism,dobakhshari2018contract}.

	% Ideally, the power system operator can subtract the baseline demand profile, which is defined as the realization of the demand if it had not responded, from the actual demand measurements to calculate the change of demand. For example, if the total demand of a commercial building should have been 30~kW during a given time period, but the actual demand turned out to be 20~kW because the building manager turned-off some loads in response to a response instructed issued by the local utility, then the effective demand response is 10~kW. As shown in Fig.\ref{fig:dr}, system operators can obtain the effective demand response components by trying to accurately forecast the baseline demand, and then subtract it from the actual net demand measurements. 
	
	\subsection{Demand response models}
	% The purpose of demand response model is to predict how will a participant respond to a given tariff. Fig.~\ref{fig:dr} shows an example of how a consumer deviates from its baseline demand in response to a time-of-use tariff, in which the consumer increased its demand during off-peak periods and decreased demand during peak periods, such as shifting the use of water heaters, to reduce its utility bill, and the utility achieves peak shaving through this tariff~\cite{pinson2014benefits,aalami2010demand}.
	
	The decision-making process of an incentive-based DR participant can be generalized as an optimization problem with corresponding objective functions and constraints. The goal of the optimization is to minimize the total cost of the consumer responding to DR incentives which include the DR reward and the disutility. Quadratic objective functions are often used for modeling disutilities based on the intuition that the marginal ``sacrifice'' increases as a consumer reduces more demand, such as turning off lighting loads or reducing the HVAC comfort level~\cite{hassan2020hierarchical, hug2015consensus, papavasiliou2010market}. The approach of representing DR decision-making with a disutility function has been widely adopted in DR coordination studies to design incentives that properly navigate consumers' behavior~\cite{paudel2020peer,tushar2014three}.
	
	Constraints in the demand response model reflect the limits that a consumer can change its electricity demand regardless of the disutility cost. This could include demand shifting needs to shift the electricity consumption from peak price periods to off-peak periods,  including rescheduling laundries, temporarily lower water heater settings, or use behind-the-meter energy storage to purposefully arbitrage electricity price differences. Operation of behind-the-meter energy storage and electric vehicles, subject to more sophisticated state-of-charge evolution and efficiency constraints, can also be represented as linear constraints~\cite{ziras2021prosumer, xu2020lagrangian}.

	\section{Modeling}
	In this section, we present the end-to-end learning framework. 
	We first describe the DR agent model as a constrained quadratic optimization problem. With such a definition, we model the demand response agent identification problem as a joint learning problem, to identify both the baseline demand and the unknown parameters in the agent model. 
	
	\textbf{Note on vector representation.} For the convenience of mathematical presentation, $D_i$, $y_i$, $z_i$ and $\lambda_i$, are all vectors with the size same as the considered agent optimization time horizon $T$ instead of a scalar. We will add the subscript $t$ to represent a scalar value for a specific time period $t$, hence $y_{i} = \{y_{i, t}|t\in\{1,2,\dotsc, T\}\}$ means the demand response vector with $T$ dimensions. Sometimes, we drop the index $i$ and just use $D_t, y_t, z_t, \lambda_t$ to denote the $t$-th element of the corresponding vector. In addition,
	$x$ is the input features for baseline demand forecast (eg. temperature, calendar information and etc), and 
	$\beta$, $\theta$ are the model parameters to be learnt. 
	
	\subsection{Demand response agent model}
	We assume a demand response agent conducts a private optimization over time horizon $T$ to determine its responses $\{y_t\}_{t=1}^{T}$ to a time-of-use price $\{\lambda_t\}_{t=1}^{T}$ on top of its baseline demand $\{D_t\}_{t=1}^{T}$. The objective of the demand response agent is to minimize its total electricity bill and the personal discomfort as follows,
	\begin{subequations}\label{eq:agent_model}
		\begin{align}
		\textstyle  \min_{y}\; g(D_t, y_t; \lambda_t, \theta) := & \sum_{t=1}^T \lambda_t y_t + \frac{\alpha(D_t)}{2} y_t^2 \,,\label{eq:agent_obj}\\
		\text{ s.t. } & \underline{P} \leq y_t \leq \overline{P}\,, \forall t \quad : (\muu_t, \muo_t)\label{eq:agent_contr1}\\
		& \underline{E} \leq \sum_{\tau=1}^{t} y_{\tau} \leq \overline{E}\,,\forall t \quad : (\nuu_t, \nuo_t) \label{eq:agent_contr2}
		\end{align}
	\end{subequations}
	where $y_t$ is the optimization decision variable, and $\theta = \{\alpha, \underline{P}, \overline{P}, \underline{E}, \overline{E} \}$ includes parameters in the agent model. We model the personal discomfort as a quadratic function with respect to the preferred consumption schedule (no demand response)~\cite{jacquot2017demand} and $\{\alpha(D_t)\}_{t=1}^{T}$ is a time-varying comfort deviation coefficient. We model $\alpha$ a function of $D_t$ indicating the user comfort deviation coefficient is dependent on the baseline demand. For example, reducing the same amount of demand may incur a higher discomfort to a user during low demand periods.
	%\footnote{The comfort deviation coefficient $\alpha_t$ can also depend on baseline demand $D_t$, which we discuss later in Remark 2.}. 
	\eqref{eq:agent_contr1} models the response limit at each time period in which $\overline{P}$ is the maximum demand increase and $\underline{P}$ is the maximum decrease. \eqref{eq:agent_contr2} is the demand response balance constraint capping the total amount of demand response over the considered time periods between $\underline{E}$ and $\overline{E}$. This constraint loosely models resources such as thermal balancing constraint or energy storage state-of-charge constraints~\cite{mathieu2014arbitraging}. We also list the dual variables associated with these constraints $\{\muu_t, \muo_t,\nuu_t, \nuo_t\}$, these dual variables will later be utilized for designing our gradient-based training algorithm.
	
	In our DR identification problem, we assume parameters in $\theta$ are \emph{unknown} to the system operator. For example, the system operator does not know what are the discomfort coefficient for their DR participants (i.e., $\alpha$), nor the DR response capacity that a participant can provide (i.e., $\{\underline{P}, \overline{P}, \underline{E}, \overline{E}\}$).

	\subsection{Baseline demand forecast model}
	\label{subsect:baseline}
	Baseline demand forecast is an important topic in power system planning and operation, and there have been many works proposed for both individual consumer's and aggregated users' load forecasting, such as linear models~\cite{li2017sparse}, deep learning methods~\cite{shi2017deep,wang2019probabilistic} and many more. Our framework is flexible to incorporate different baseline demand forecast methods as a plug-in module, as long as it supports the automatic differentiation~\cite{paszke2017automatic}. 
	
	In particular, we use a fully connected multi-layer perceptron (MLP) model for the baseline demand forecast in all our experiments. Inputs to the module include useful features such as weather data (eg. temperature, humidity, wind speed), calendar information (e.g., month, day, weekday/weekend), and effects of human activities. The baseline demand forecast model is represented with the following formulation,
	\begin{align}
	\hat{D} = f(x;\beta)\label{eq:bf}
	\end{align}
	where $\hat{D} = \{\hat{D_t}\}_{t=1}^{T} \in \mathbb{R}^T$ is the predicted baseline demand, $x \in \mathbb{R}^{m}$ is the demand forecast features and $f(\cdot)$ is the baseline demand prediction network with model parameters $\beta$. 
	
	\subsection{Joint baseline grounding and agent model identification}
	The goal of the system operator is to jointly identify the DR agent model $\theta = \{\alpha, \underline{P}, \overline{P}, \underline{E}, \overline{E}\}$ and baseline demand $\{D_t\}_{t=1}^{T}$. However, the challenge is that the system operator cannot directly observe agent's response $\{y_t\}_{t=1}^{T}$ with respect to the incentive price signal $\{\lambda_t\}_{t=1}^{T}$. Instead, the system operator only observes agent's \emph{net demand} $\{D_t+y_t\}_{t=1}^{T}$, which is the summation of agent's baseline demand and demand response. Existing works usually take a two-step approach. In particular, it first estimates the baseline demand using historical demand data, then subtracts baseline estimation from observed net demand for agent DR model identification. Yet, as we have discussed in the literature review,  in this approach the system operator must have access to a sufficient amount of ``vanilla'' demand data to make sure the baseline estimation is fair and accurate. However, such data might not be available if the participant has been constantly responding to DR signals. 
	
\begin{figure}[htbp]
		\centering
		\includegraphics[width=3.2in]{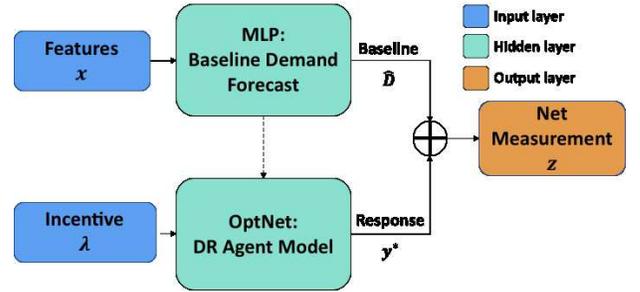}
		\caption{Illustration of the proposed end-to-end joint DR agent identification and baseline demand estimation framework.} 
		\label{fig:nn1}
	\end{figure}

	Different from the two-step approach, we propose an end-to-end differential learning approach to \emph{jointly} minimize the baseline prediction and agent model identification errors with a set of historical net demand measurements consisting $N$ days or scenarios, and each scenario includes $T$ time steps. In this case, the system operator solves the following problem
	\begin{subequations}\label{eq:TM3}
		\begin{align}
		\min_{\theta, \beta}\; L := &\frac{1}{2N} \sum_{i=1}^N \sum_{t=1}^T \Big(\hat{D}_{i, t} + y^{*}_{i, t}-z_{i, t}\Big)^{2}\label{eq:training_obj}\\
		\text{s.t. } & \{\hat{D}_{i,t}\}_{t=1}^T = f(x_i;\beta) \\
		& \{y^{*}_{i,t}\}_{t=1}^T \in \arg\min_{y}\; g(\{\hat{D}_{i,t}, y_{i, t}\}_{t=1}^T; \{\lambda_{i, t}\}_{t=1}^{T}, \theta) \nonumber\\
		&\qquad\quad \text{s.t. \eqref{eq:agent_contr1} and \eqref{eq:agent_contr2}}
		\end{align}
	\end{subequations}
	in which the training~/~identification inputs include
	\begin{itemize}
		\item $z_{i,t}$ is the net demand measurement over scenario $i$ and time period $t$;
		\item $x_i$ is the baseline demand prediction features;
		\item $\lambda_{i,t}$  is the demand response incentives over scenario $i$ and time period $t$;
		\item $f(\cdot)$ is the MLP model used for baseline demand prediction;
		\item $g(\cdot)$ is the agent demand response model.
	\end{itemize}
	and the problem has the following variables
	\begin{itemize}
		\item $\theta$ is the DR agent model parameters;
		\item $\beta$ is the MLP network weights used for the baseline demand model;
		\item $\hat{D}_{i,t}$ is the baseline demand over scenario $i$ and time period $t$;
		\item $y^{*}_{i, t}$ is the DR agent response decision over scenario $i$ and time period $t$.
	\end{itemize}
	The joint identification problem minimizes the mean squared loss between the predicted net demand $(\hat{D}_{i,t} + y^{*}_{i,t})$ and the observed net demand $z_{i,t}$ for $N$ training samples over $T$ horizon. 
	%Loss function $L$ measures the distance between the predicted and observed net demand. For instance, one common loss function is the L2 distance, where $L(\hat{D}_i + y^{*}_i; z_i) = ||\hat{D}_i + y^{*}_i-z_i||^{2} = \sum_{t=1}^{T} (\hat{D}_{i, t} + y^{*}_{i, t}-z_{i, t})^{2}$. 
	Fig~\ref{fig:nn1} visualizes the proposed end-to-end joint baseline demand estimation and agent model identification framework.
	Specifically, the top green colored block in Fig~\ref{fig:nn1} is the baseline demand forecast module that maps the demand forecast features to baseline demand prediction; and the bottom green colored block is the demand response forecast, that takes DR signals and predict the DR agent response. The summation of these two intermediate predictions construct the net demand prediction $(\hat{D}_i + y^{*}_i)$, which is compared against the true net demand data $z_i$ as the model loss function~\eqref{eq:training_obj}. 

	\section{Training Method}
	\label{subsect:dr_optnet}
	We propose a gradient-descent approach to solve the joint baseline and DR agent identification problem \eqref{eq:TM3}. We start by taking the gradient of the training objective function $L$ with respect to model parameters $\theta$ and $\beta$
	% \begin{subequations}
	% \begin{align}
	%     \frac{\partial L}{\partial \theta} &= \frac{1}{NT} \sum_{i=1}^N \sum_{t=1}^T \Big(\hat{D}_{i, t} + y^{*}_{i, t}-z_{i, t}\Big)\Big(\frac{\partial \hat{D}_t}{\partial \theta} + \frac{\partial y^*_t}{\partial \theta}\Big) \\
	%     \frac{\partial L}{\partial \beta} &= \sum_{i=1}^N \sum_{t=1}^T \Big(\hat{D}_{i, t} + y^{*}_{i, t}-z_{i, t}\Big)\Big(\frac{\partial \hat{D}_t}{\partial \beta} + \frac{\partial y^*_t}{\partial \beta}\Big)
	% \end{align}
	% \end{subequations}
	% and now we are interested in calculating the four partial derivative terms $\frac{\partial \hat{D}_t}{\partial \theta}, \frac{\partial y^*_t}{\partial \theta}, \frac{\partial \hat{D}_t}{\partial \beta}, \frac{\partial y^*_t}{\partial \beta}$. First note that gradients of $\frac{\partial \hat{D}_t}{\partial \theta}=0$ and $\frac{\partial \hat{D}_t}{\partial \beta}$ is the gradient of the MLP, which can be easily computed via backpropagation. Calculating the gradients of the DR response $y^{*}_{i, t}$ is more sophisticated and is the focus of our proposed method. In this section, we will first explain how to compute $\frac{\partial y^*_t}{\partial \theta}$ using the OptNet architecture~\cite{amos2017optnet}, then we calculate $\frac{\partial y^*_t}{\partial \beta}$ using the chain rules as $\frac{\partial y^*_t}{\partial \beta} = \frac{\partial y^*_t}{\partial D_t} \frac{\partial D_t}{\partial \beta}$.
	\begin{subequations}
		\begin{align}
		\frac{\partial L}{\partial \theta} &= \frac{1}{N} \sum_{i=1}^N \sum_{t=1}^T \Big(\hat{D}_{i, t} + y^{*}_{i, t}-z_{i, t}\Big)\Big(\frac{\partial y^*_t}{\partial \theta}\Big) \\
		\frac{\partial L}{\partial \beta} &= \frac{1}{N} \sum_{i=1}^N \sum_{t=1}^T \Big(\hat{D}_{i, t} + y^{*}_{i, t}-z_{i, t}\Big)\Big(\frac{\partial \hat{D}_t}{\partial \beta} + \frac{\partial y^*_t}{\partial \beta} \Big)
		\end{align}
	\end{subequations}
	and now we are interested in calculating the three partial derivative terms $\frac{\partial y^*_t}{\partial \theta}, \frac{\partial \hat{D}_t}{\partial \beta}, \frac{\partial y^*_t}{\partial \beta}$. The demand response $y^*_t$ is dependent on the baseline demand coefficients $\beta$ because the discomfort coefficient $\alpha$ is dependent on the baseline $D_t$ as shown in \eqref{eq:agent_model}. Also note that gradient $\frac{\partial \hat{D}_t}{\partial \beta}$ is the gradient of the MLP, which can be easily computed via backpropagation. Calculating the gradients of the DR response $y^{*}_{i, t}$ is more sophisticated and is the focus of our proposed method. 
	
	\subsection{Gradients of the DR response}
	\label{subsect:dr_optnet}
	We use the OptNet architecture~\cite{amos2017optnet} to represent the agent identification problem \eqref{eq:agent_model}. The OptNet architecture leverages prior knowledge about the optimization and thereby, favouring sample efficiency. To obtain the demand response forecast, it solves the optimization problem in the forward loop, with the learnt parameters. Then it compares the difference between the optima of the optimization problem with the learnt parameters, and the optima of the true optimization problem, and use the difference as the loss function to update the unknown optimization parameters.
	%In particular, OptNet encodes the agent utility optimization problem with its flexibility constraint as a neural network, where the unknown optimization parameters  are weights of the layer to be learnt. 
	
	Our objective is to calculate $\frac{\partial y^*_t}{\partial \theta}$ and to backpropagate the training loss with respect to the agent model parameters $\theta$. To this end, we ignore the dependency of $\alpha$ over $D_t$ for now and treat it as a parameter instead of a function. In the next subsection on end-to-end training, we will discuss incorporate the dependency $\alpha$ over $D_t$ using derivative chain rules. 
	We first converting the agent model into a set of equations using the Karush–Kuhn–Tucker (KKT) conditions. The KKT conditions of optimization problem~\eqref{eq:agent_model} can be written as, for $t = 1, ..., T$,
	\begin{subequations}\label{eq:kkt}
		\begin{align}
		\textstyle \lambda_t + \alpha y_t^{*} + \muo_t^{*} -\muu_t^{*} + \sum_{\tau=t}^T \nuo_{\tau}^{*} -  \sum_{\tau=t}^T \nuu_{\tau}^{*} & = 0 \label{eq:kkt1}\\
		\muo_t^{*} (y_t^{*} - \overline{P})   = 0\,,\quad \muo_t^{*} &\geq 0 \label{eq:kkt2}\\
		\muu_t^{*}  (\underline{P}-y_t^{*}) = 0\,,\quad \muu_t^{*} &\geq 0 \label{eq:kkt3}\\
		\textstyle \nuo_t^{*} (\sum_{\tau=1}^t y_{\tau}^{*} - \overline{E})    = 0\,,\quad \nuo_t^{*} &\geq 0 \label{eq:kkt4}\\
		\textstyle \nuu_t^{*} (\underline{E} - \sum_{\tau=1}^t y_{\tau}^{*})   = 0\,,\quad \nuu_t^{*} &\geq 0  \label{eq:kkt5}
		\end{align}
	\end{subequations}
	where \eqref{eq:kkt1} is the stationarity condition of the DR agent problem \eqref{eq:agent_model}, \eqref{eq:kkt2}--\eqref{eq:kkt5} are the complementary slackness conditions associated with constraints \eqref{eq:agent_contr1} and \eqref{eq:agent_contr2}. $y^{*}, \underline{\mu}^{*}, \overline{\mu}^{*}, \underline{\nu}^{*}, \overline{\nu}^{*}$ are the optimal primal and dual variables. 
	% where $y^{*}, \nu^{*}$ and $\mu^{*}$ are the optimal primal and dual variables, and $\circ$ is the element-wise product operator. The three equations represent the stationarity, primal feasibility and complementary slackness conditions for the optimal solution.
	
	Next, we calculate the derivative of the optimal response $y_t^{*}$ with respect to model parameters $\theta = \{\alpha, \underline{P}, \overline{P}, \underline{E}, \overline{E}\}$. We first take the total derivatives of the KKT conditions and summarize in compact matrix form, which is given in Eq~\eqref{eq:total_derivative_kkt_compact},
	\begin{strip}
		\begin{align}\label{eq:total_derivative_kkt_compact}
		&\underbrace{\begin{bmatrix}
			\Lambda(\mathbbm{1}\alpha) & I & -I & \Gamma^\top & -\Gamma^\top\\
			\Lambda(\overline{\mu}^{*}) & \Lambda(y^{*}\!-\!\mathbbm{1}\overline{P}) & 0 & 0 & 0\\
			-\Lambda(\underline{\mu}^{*}) & 0 & \Lambda(\mathbbm{1} \underline{P}\!-\! y^{*}) & 0 & 0\\
			\Lambda(\overline{\nu}^{*}) \Gamma & 0 & 0 & \Lambda(\Gamma y^{*}\!-\! \mathbbm{1}\overline{E}) & 0\\
			-\Lambda(\underline{\nu}^{*}) \Gamma & 0 & 0 & 0 & \Lambda(\mathbbm{1}\underline{E}\!-\!\Gamma y^{*})
			\end{bmatrix}}_{A} % \nonumber\\
		* \begin{bmatrix}
		dy\\
		d\overline{\mu}\\
		d\underline{\mu}\\
		d\overline{\nu}\\
		d\underline{\nu}
		\end{bmatrix} = 
		-\begin{bmatrix}\Lambda(y^{*}) \mathbbm{1} d\alpha \\
		-\Lambda(\overline{\mu}^{*}) \mathbbm{1} d\overline{P}\\
		\Lambda(\underline{\mu}^{*}) \mathbbm{1} d\underline{P}\\
		-\Lambda(\overline{\nu}^{*}) \mathbbm{1} d\overline{E}\\
		\Lambda(\underline{\nu}^{*}) \mathbbm{1} d\underline{E}
		\end{bmatrix}
		\end{align}
	\end{strip}
	where $\mathbbm{1}$ is an all-one vector and $\Gamma$ is a lower triangular matrix with appropriate dimension, and $\Lambda(\cdot)$ creates a diagonal matrix from a vector. The deviation process of Eq~\eqref{eq:total_derivative_kkt_compact} is provided in the Appendix. 
	
	Our goal is to obtain the derivative of the optimal primal solution $y^{*}$ with respect to the optimization parameters $\theta = \{\alpha, \underline{P}, \overline{P}, \underline{E}, \overline{E} \}$ (we can also compute the Jacobian of optimal dual solutions with respect to the optimization parameters using the same approach). We divide both sides of \eqref{eq:total_derivative_kkt_compact} with the parameter derivatives $d\alpha, d\underline{P}, d\overline{P}, d\underline{E}, d\overline{E}$ respectively, and multiply both sides with $A^{-1}$:
	\begin{subequations}\label{eq:jac}
		\begin{align}
		\begin{bmatrix}
		\frac{dy}{d\alpha}\;
		\frac{d\overline{\mu}}{d\alpha}\;
		\frac{d\underline{\mu}}{d\alpha}\;
		\frac{d\overline{\nu}}{d\alpha}\;
		\frac{d\underline{\nu}}{d\alpha}
		\end{bmatrix}^\top &= - A^{-1}\begin{bmatrix}y^{*} \;
		\vec{0}\;
		\vec{0}\;
		\vec{0}\;
		\vec{0}
		\end{bmatrix}^\top \label{eq:jac1}\\
		\begin{bmatrix}
		\frac{dy}{d\overline{P}}\;
		\frac{d\overline{\mu}}{d\overline{P}}\;
		\frac{d\underline{\mu}}{d\overline{P}}\;
		\frac{d\overline{\nu}}{d\overline{P}}\;
		\frac{d\underline{\nu}}{d\overline{P}}
		\end{bmatrix}^\top &= -A^{-1}\begin{bmatrix}\vec{0} \;
		-\overline{\mu}^{*} \;
		\vec{0}\;
		\vec{0}\;
		\vec{0}
		\end{bmatrix}^\top \label{eq:jac2}\\
		\begin{bmatrix}
		\frac{dy}{d\underline{P}}\;
		\frac{d\overline{\mu}}{d\underline{P}}\;
		\frac{d\underline{\mu}}{d\underline{P}}\;
		\frac{d\overline{\nu}}{d\underline{P}}\;
		\frac{d\underline{\nu}}{d\underline{P}}
		\end{bmatrix}^\top &= -A^{-1}\begin{bmatrix}\vec{0} \;
		\vec{0}\;
		\underline{\mu}^{*} \;
		\vec{0}\;
		\vec{0}
		\end{bmatrix}^\top \label{eq:jac3}\\
		\begin{bmatrix}
		\frac{dy}{d\overline{E}}\;
		\frac{d\overline{\mu}}{d\overline{E}}\;
		\frac{d\underline{\mu}}{d\overline{E}}\;
		\frac{d\overline{\nu}}{d\overline{E}}\;
		\frac{d\underline{\nu}}{d\overline{E}}
		\end{bmatrix}^\top &= -A^{-1}\begin{bmatrix}\vec{0}\;
		\vec{0}\;
		\vec{0}\;
		-\overline{\nu}^{*} \;
		\vec{0}
		\end{bmatrix}^\top \label{eq:jac4}\\
		\begin{bmatrix}
		\frac{dy}{d\underline{E}}\;
		\frac{d\overline{\mu}}{d\underline{E}}\;
		\frac{d\underline{\mu}}{d\underline{E}}\;
		\frac{d\overline{\nu}}{d\underline{E}}\;
		\frac{d\underline{\nu}}{d\underline{E}}
		\end{bmatrix}^\top &= -A^{-1}\begin{bmatrix}\vec{0} \;
		\vec{0}\;
		\vec{0}\;
		\vec{0}\;
		\underline{\nu}^{*} 
		\end{bmatrix}^\top \label{eq:jac5}
		\end{align}
	\end{subequations}
	Note that the derivative of a parameter with respect to itself (for example,  $\frac{d\alpha}{d\alpha}$) is one while the derivative to different parameters (for example, $\frac{d\overline{P}}{d\alpha}$) are zeros.
	Therefore we obtain $\frac{\partial y^{*}}{\partial \alpha}, \frac{\partial y^{*}}{\partial \overline{P}}, \frac{\partial y^{*}}{\partial \underline{P}}, \frac{\partial y^{*}}{\partial \overline{E}}, \frac{\partial y^{*}}{\partial \underline{E}}$ each as the first element of \eqref{eq:jac1}--\eqref{eq:jac5}.

	% For example, if we want to compute the Jacobian of $\frac{\partial y^{*}}{\partial \alpha} \in \mathbb{R}^{T \times T}$, we would divide $d \alpha$ from both side,
	% \begin{equation*}
	%     A \begin{bmatrix}
	%     \frac{dy}{d\alpha}&
	%     \frac{d\overline{\mu}}{d\alpha}&
	%     \frac{d\underline{\mu}}{d\alpha}&
	%     \frac{d\overline{\nu}}{d\alpha}&
	%     \frac{d\underline{\nu}}{d\alpha}
	% \end{bmatrix}^\top = -\begin{bmatrix}\Lambda(y^{*}) &
	%     0&
	%     0&
	%     0&
	%     0
	%     \end{bmatrix}^\top
	% \end{equation*}
	% Solving the above linear system, the resulting value of $\frac{dy}{d\alpha}$ would be the desired Jacobian. Similar approach applies to calculating the other Jacobian terms. 
	\vspace{6pt}
	\textbf{Remark 1.} An implicit assumption for the above derivation regarding the Jacobian terms $\frac{\partial y^{*}}{\partial \alpha}, \frac{\partial y^{*}}{\partial \overline{P}}, \frac{\partial y^{*}}{\partial \underline{P}}, \frac{\partial y^{*}}{\partial \overline{E}}, \frac{\partial y^{*}}{\partial \underline{E}}$ is that the optimal solution is continuously differentiable in a neighborhood of the optimization parameters. However, for a general convex optimization problem, the optimal solution sets are not necessarily continuously differentiable in the optimization parameters. Paper~\cite{barratt2018differentiability} provided sufficient conditions under which these Jacobian are well-defined. We consider the DR agent optimization model as a Quadratic Program (QP), which satisfies Assumptions 1-3 in~\cite{barratt2018differentiability}, thus the Jacobians are well-defined. Beyond the quadratic form DR agent model, our framework can also be used in general convex DR agent models, where both the objective function and constraints are twice continuously differentiable in the optimization variable $y$, and continuously differentiable in the optimization parameters.
	
	\subsection{End-to-end training} \label{sec:e2e}
	Remember that we consider the problem setting where the system operator only has access to historical demand response incentive signal $\lambda_i$, baseline demand feature $x_i$ and the net demand measurement $z_i$ (baseline demand + demand response). The system operator solves the joint learning problem as stated in \eqref{eq:TM3}, in order to jointly identify the baseline demand and agent DR model. To this end, we propose an end-to-end (E2E) training framework that combines both the gradient of the baseline demand forecast module and the demand response agent module. The model parameters are updated as follows during iteration $k$:
	\begin{subequations}
		\begin{align}
		\text{Baseline: }\beta^{(k+1)} &\leftarrow \beta^{(k)} + \eta_1 \Big(\frac{\partial L}{\partial D} + \frac{\partial L}{\partial y^{*}} \frac{\partial y^{*}}{\partial \alpha} \frac{\partial \alpha}{\partial D}\Big)\frac{\partial D}{\partial \beta} \,, \label{weight_f1}\\
		\text{DR agent: }\theta^{(k+1)} &\leftarrow \theta^{(k)} + \eta_2 \frac{\partial L}{\partial y^{*}} \frac{\partial {y}^{*}}{\partial \theta} \,, \label{weight_f2}
		\end{align}
	\end{subequations}
	where $\eta_1$ and $\eta_2$ are learning rates for baseline and DR agent modules. $\frac{\partial L}{\partial D}$ and $\frac{\partial L}{\partial y^{*}}$ are gradients of the loss function with respect to the intermediate baseline demand and agent response predictions. $\frac{\partial D}{\partial \beta}$ is the gradient of $D$ with respect to the baseline demand forecast model weights. For the MLP network model used in this paper, we compute the gradient via backpropagation provided by PyTorch. The gradient update of the baseline module also includes $\frac{\partial y^{*}}{\partial \alpha} \frac{\partial \alpha}{\partial D}$ as we assume the DR agent's discomfort coefficient $\alpha$ depends on the baseline demand. $\frac{\partial y^{*}}{\partial \alpha}$ is calculated in \eqref{eq:jac1}, and $\frac{\partial \alpha}{\partial D}$ depends on the agent model. 
	Gradient updates of the DR agent module is solely based on
	$\frac{\partial y^{*}}{\partial \theta}$ which we calculate according to \eqref{eq:jac}. Notably, this training step has a separable structure that the demand forecast neural network module has no common parameters with the differentiable optimization module, hence their gradient will not impact each other.
	
	Compared to classic two-step methods that first predict the baseline demand and then estimate the agent DR model and response amount, our end-to-end framework alleviates error accumulation since both modules are optimized simultaneously. Compared with a fully connected network overall features, our design embeds prior knowledge about the agent behavior (i.e., the utility optimization problem), which reduces the sample and computational complexity in magnitudes, and also providing better model explainability. 
	
	The training process is summarized as following for a given training data set $\{x_{i}, \lambda_{i}, z_{i}\}$
	\begin{enumerate}
		\item Initialize unknown parameters for the baseline forecast module $\beta^{(1)}$ and the DR agent module $\theta^{(1)}$; set the iteration number to 1.
		\item At $k$th iteration
		\begin{enumerate}
			\item Forward the baseline demand prediction module \eqref{eq:bf} with feature input $x_{i}$ and parameters $\beta^{(k)}$, record the baseline prediction result  $\hat{D}_{i}^{(k)}$.
			\item Forward solve the DR problem \eqref{eq:agent_model} with price $\lambda_{i}$ and parameters $\theta^{(k)}$ using an optimization solver, record the primal results ${y_i^*}^{(k)}$ and the dual results ${\muu_i^*}{(k)}, {\muo_i^*}{(k)},{\nuu_i^*}{(k)}, {\nuo_i^*}{(k)}$.
			\item Add up the two prediction as the net demand prediction $\hat{D}_{i}^{(k)}+{y_i^*}^{(k)}$, compare it against the observed net demand $z_i$ and obtain loss $L= ||\hat{D}_{i}^{(k)}+{y_i^*}^{(k)}-z_i||_2^2$.
			\item Update $\beta^{(k+1)}$ and $\theta^{(k+1)}$ with Eq~\eqref{weight_f1} and~\eqref{weight_f2}. Proceed to the next iteration.
		\end{enumerate}
	\end{enumerate}
	
	\textbf{Remark 2. Warm start.} In experiments, we observed that adding a warm start to the baseline demand forecast module using net measurements could improve the performance. In particular, the warm start is stated as
	\begin{align}\label{eq:WS}
	\min_{\beta}\; &\textstyle \frac{1}{N} \sum_{i=1}^{N} ||\hat{D}_i - z_i||^2
	\text{ s.t. } \hat{D}_i = f(x_i;\beta) 
	\end{align}
	We note that the warm start trains the baseline demand prediction model using net demand measurements $z_i$ \emph{rather than} historical baseline demand (which the system operator may not have access to). The optimized model parameter $\beta$ is then used as a warm starting point for the end-to-end training stage, in which $\beta$ is optimized together with DR agent model parameters $\theta$ following Eq~\eqref{weight_f1} and~\eqref{weight_f2}.
	
	\textbf{Remark 3. Demand response dependency over baseline.} We ends this section by discussing how the proposed framework handles the dependency of demand response over the baseline demand. In some demand response scenarios like load shedding and peak shaving, the demand response is dependent over the baseline demand. In other scenarios like behind-the-meter energy storage price response, the change of demand may not depend on the baseline demand. This difference is modeled as $\frac{\partial \alpha}{\partial D}$ that if the disutility parameter $\alpha$ is dependent/independent of the baseline demand estimation $D$. If there is no dependency, we can set this value to zero and then the baseline gradient update reduces to $\beta^{(k+1)} \leftarrow \beta^{(k)} + \eta_1 \frac{\partial L}{\partial D} \frac{\partial D}{\partial \beta}$.
	\section{Experiment Result and Discussion}
	We demonstrate the performance of the proposed framework in identifying the DR agent model and baseline demand with both synthetic dataset and real world demand response dataset. In the synthetic dataset experiment, we compared the identification result with the ground-truth DR model, while with the real world dataset we compared end-to-end training with a benchmark baseline estimation method using a control group. We use Pytorch to build the models and run the training process in Google Colab with GPU acceleration.
	
	\subsection{Demand response identification with synthetic dataset}
	\label{sec:result2}
	
	We generate DR data using real demand profiles as the baseline, with added response to price signals. For the baseline demand, we assume a demand response aggregator works directly with aggregated measurement which is viewed as one response agent. We use smart meter data from the Iowa Distribution Test Systems Dataset~\cite{bu2019time}. The system consists of 3 feeders, 240 nodes, and 1120 customers, all of which are equipped with smart meters. These smart meters measure hourly energy consumption (kWh) in the year 2017. The electricity consumption data is at the secondary distribution transformer level. In particular, we use the aggregated demand data from Feeder 1, Bus 1004 for the experiment. We also include the temperature data of Columbus from the Climate Data Online (CDO) portal in the National Climatic Data Center (NCDC) website \cite{noaa} as baseline load forecast features. We use 200 days of data for training and 60 days for testing the demand response prediction results using the identified parameters. 
	Table~\ref{table_stats} provides the detailed statistics of the demand profile.
	\begin{table}[htbp]
		\renewcommand{\arraystretch}{1.5}
		\centering
		\caption{Summary of Statistics of the Iowa Distribution Bus 1004}
		\begin{tabular}{cccc}
			\hline
			\hline
			Max (kW) & Min (kW) & Mean (kW) & Std (kW)\\
			\hline
			41.534 & 2.401 & 11.332 & 6.166 \\
			\hline
			\hline
		\end{tabular}
		\label{table_stats}
	\end{table}
	
	\begin{figure}[t]%
		\centering
		\subfloat[]{
			\includegraphics[trim = 0mm 00mm 0mm 0mm, clip, width = .95\columnwidth]{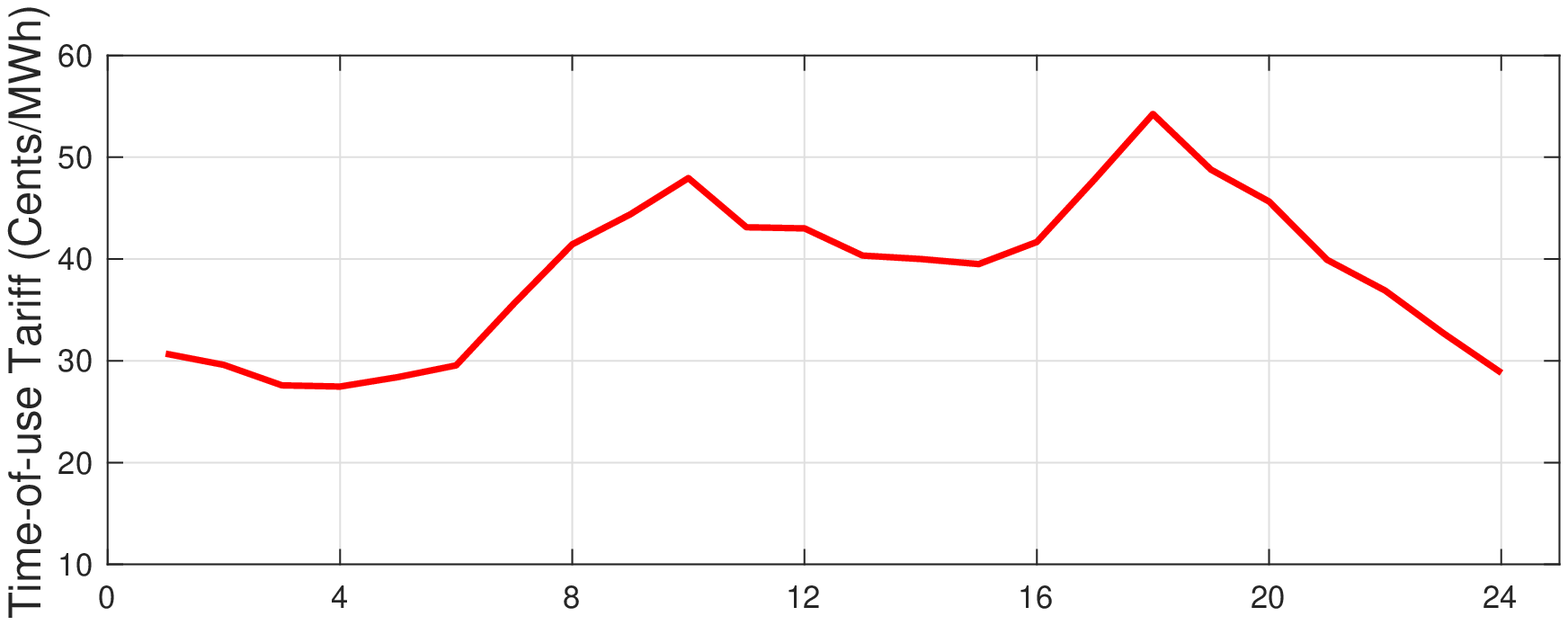}
			\label{fig:data_vis1}%
		}\\
		\subfloat[]{
			\includegraphics[trim = 0mm 00mm 0mm 0mm, clip, width = .95\columnwidth]{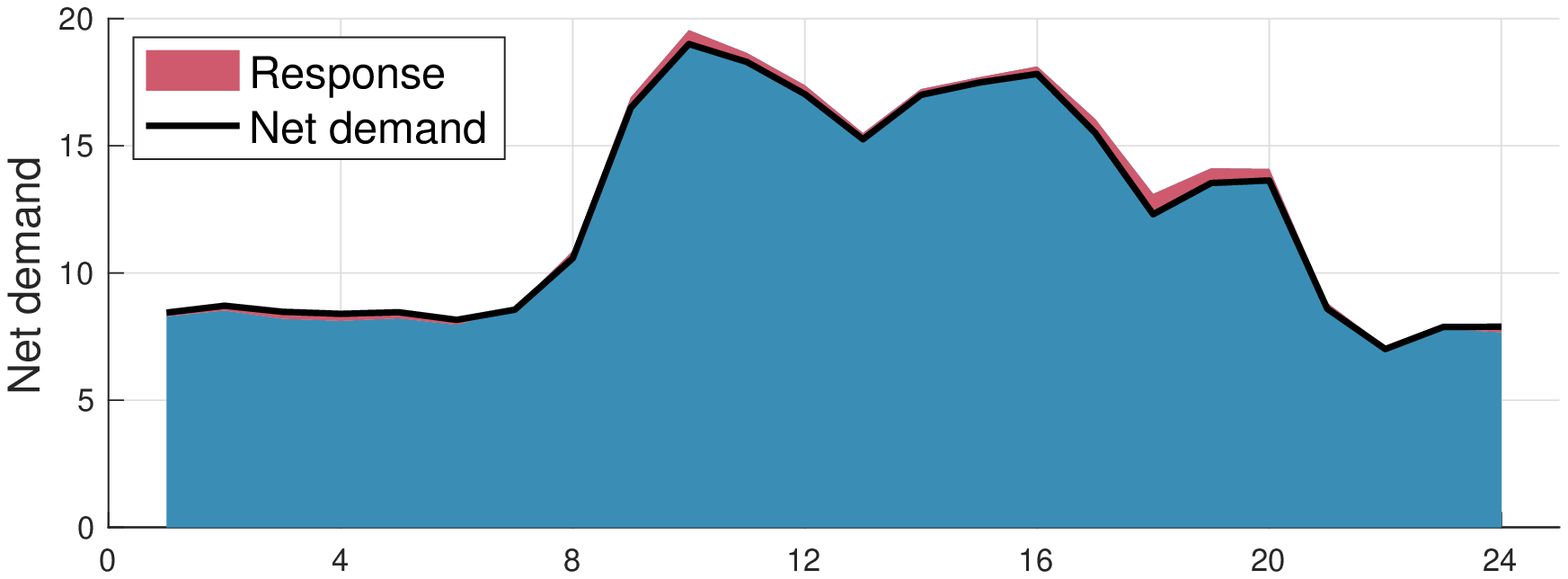}
			\label{fig:data_vis2}%
		}
		\caption{Visualization example of training and testing data (a) DR price; (b) baseline and net demand of the aggregation dataset ($\alpha = 25$, $M = 4$).}
		\label{fig:data_vis}
	\end{figure}
	
	For demand response, we consider a demand-independent demand response agent model in this case study as follows
	\begin{align}
	\textstyle  \min_{\{y_t\}} \sum_{t=1}^{T} \lambda_t y_t + \frac{\alpha}{2} y_t^2 \text{ s.t. } -M \leq \sum_{t=1}^{T} y_t \leq M %; M \geq 0
	\label{eq:demand_response}
	\end{align}
	where $\alpha$ is the comfort deviation coefficient and $M$ represents the limit of the total demand change within the optimization period. In order to test the performance of different agent models, we randomly sample the ground truth $\alpha \sim \text{Uniform}[10, 50]$ and $M \sim \text{Uniform}[1, 10]$ at each test time. $\lambda$ is the time-of-use tariff, where we use the day ahead LMP data from NYISO~\cite{nyiso}. We assume a day-ahead scheduled DR in which the optimization horizon is set to be $T = 24$ (24 hours). 
	
	We use the above agent model~\eqref{eq:demand_response} and the corresponding day ahead LMP~\cite{nyiso} to generate synthetic demand response data, for a specific sampled combination ($\alpha$, $M$). 
	The baseline demand data and the synthetic demand response data are combined to generate the net measurement data set. Fig~\ref{fig:data_vis} visualizes an example of the baseline demand and net demand for 1 day. Throughout our experiments, all MLP baseline demand forecast models are feed-forward neural networks consisting of three hidden layers with (200, 100, 100) hidden units per layer, and the demand response module uses OptNet architecture defined on optimization Eq.~\eqref{eq:demand_response}. We use Adam optimizer with a learning rate of $1e-3$ for the base load forecast module and $1e-1$ for the OptNet layer. For evaluations, we repeated for $N = 10$ times with different choices of $(\alpha, M)$ to report the average performance and the standard deviations.
	Table~\ref{table_agentid} summarizes the mean absolute error (MAE) and standard deviations for the parameter identification results. 
	\begin{table}[h]
		\renewcommand{\arraystretch}{1.5}
		\centering
		\caption{Agent Model Identification Errors (MAE).}
		\begin{tabular}{cccc}
			\hline
			\hline
			\multicolumn{2}{c}{Discomfort coefficient $\alpha$}  & \multicolumn{2}{c}{Limit of total demand change $M$}\\
			Mean  & Std &  Mean & Std \\
			\hline
			1.178 & 0.574 & 2.164 & 0.948\\
			\hline
			\hline
		\end{tabular}
		\label{table_agentid}
	\end{table}
	
\begin{figure}[htbp]%
		\centering
		\subfloat[Demand Response Prediction]{
			\includegraphics[trim = 0mm 00mm 0mm 0mm, clip, width = .9\columnwidth]{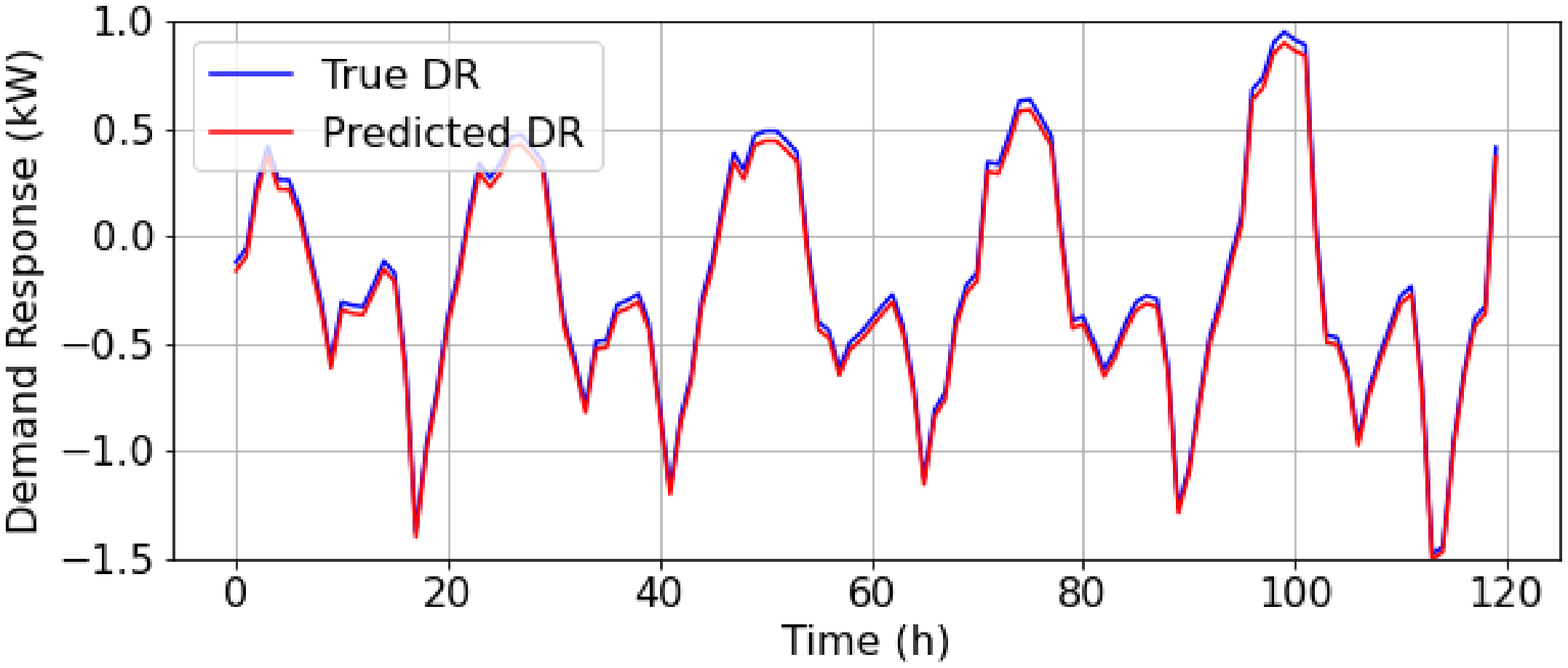}
			\label{fig:sl2}}\\
		\subfloat[A-priori Base Load Estimation]{
			\includegraphics[trim = 0mm 5mm 0mm 5mm, clip, width = .9\columnwidth]{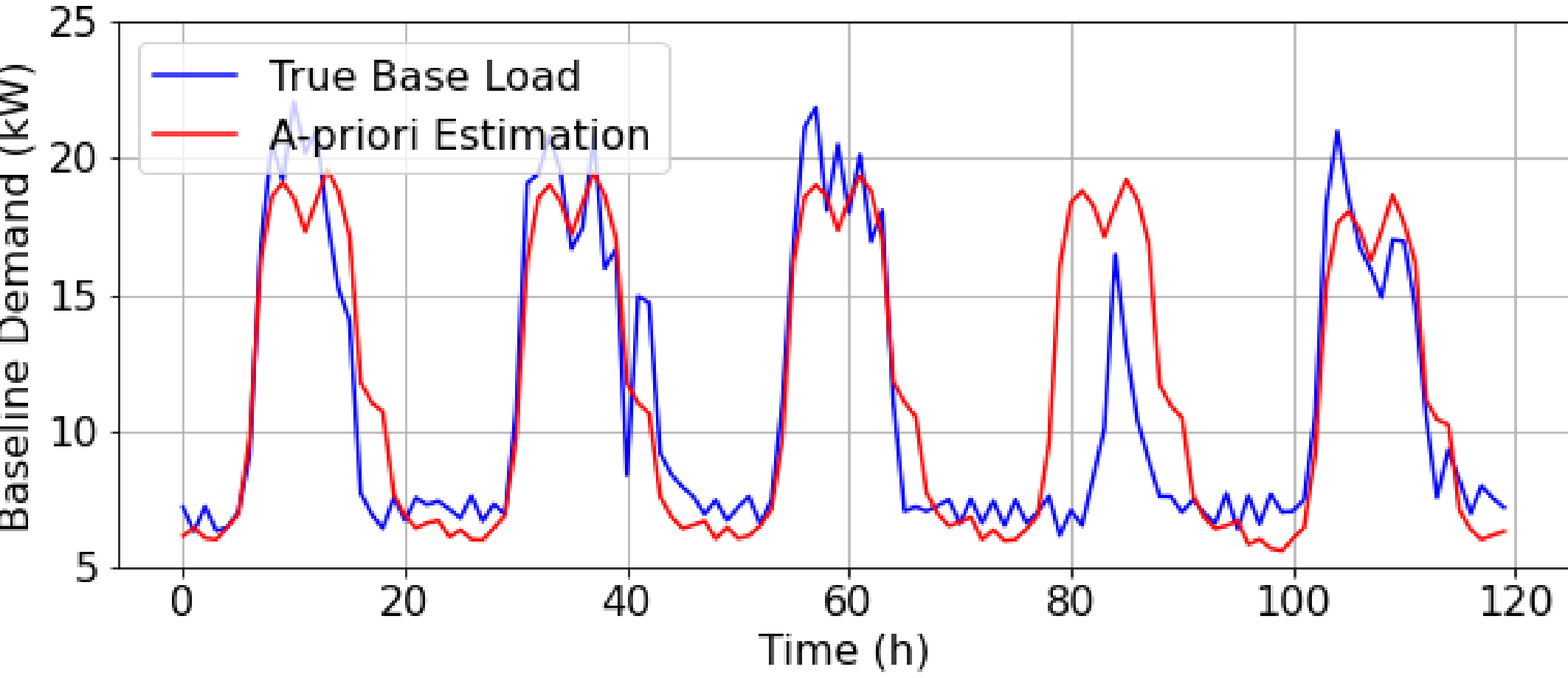}
			\label{fig:sl1}%
		}\\
		\subfloat[Ex-post Base Load Estimation]{
			\includegraphics[trim = 0mm 5mm 0mm 5mm, clip, width = .9\columnwidth]{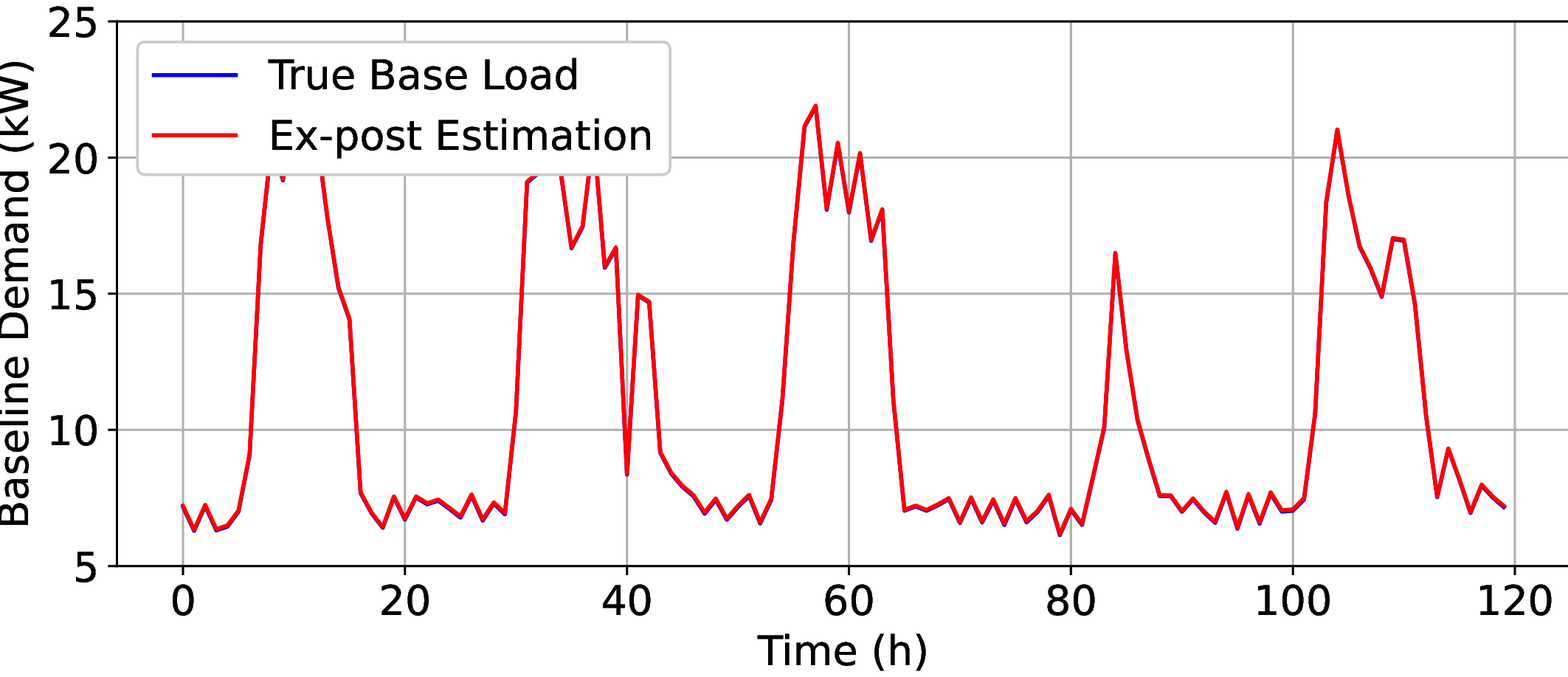}
			\label{fig:sl3}%
		}
		\caption{Visualization of DR identification and a-priori/ex-post baseline demand estimation results on 5 consecutive test day. [True M = 5.039, Predicted M= 5.957, True $\alpha= 16.447$, predicted $\alpha = 16.264$.]}
		\label{fig:prediction}
	\end{figure}

	In addition to agent demand response model identification, we compare the baseline forecast performance over two settings: \emph{a-priori baseline prediction}, in which we directly use the trained baseline forecast model to predict demand over a 24-hour horizon. A-priori baseline prediction is useful for resource scheduling and estimating feeder loads.  The other setting is \emph{ex-post baseline estimation}, in which we estimate the baseline by subtracting the demand response component (based on the identified response model parameters) from the net demand. The ex-post baseline estimation is useful to identify effective demand response and to settle payments. 
	\begin{table}[h]
		\renewcommand{\arraystretch}{1.5}
		\centering
		\caption{Baseline demand estimation error metrics (Mean absolute error (MAE)/Mean absolute Percentage Error (MAPE)).}
		\begin{tabular}{cccc}
			\hline
			\hline
			\multicolumn{2}{c}{A-priori Baseline Estimation}  & \multicolumn{2}{c}{Ex-post Baseline Estimation}\\
			MAE & MAPE & MAE & MAPE \\
			\hline
			2.93 & 27.10\% & 0.09 & 1.01\%\\
			\hline
			\hline
		\end{tabular}
		\label{table_baseline}
	\end{table}
	Results in Table~\ref{table_baseline} show a relatively large a-priori baseline prediction error due to the challenge of performing day-ahead demand prediction with limited training data and features (note that we train the model over 200 days and test over the rest 60 days). But the ex-post baseline estimation results are substantially improved. As a base for comparison, the MAE and MAPE between net demand (baseline+response) and baseline are 0.44~kW~/~4.77\% for the aggregation case. Hence, our method provides a significant improvement over ex-post baseline estimation, where the baseline estimation error is reduced by about 90\%. Finally, Fig~\ref{fig:prediction} provides an example of the end-to-end training approach performance in demand response identification, a-priori baseline estimation and ex-post baseline estimation in 5 consecutive test days. 
	% \begin{table}[htbp]
	% 	\renewcommand{\arraystretch}{1.5}
	% 	\centering
	% 	\caption{Demand baseline a-priori prediction and ex-post estimation error results (MAE/MAPE, unit: KW/\%).}
	% 	\begin{tabular}{lcccc}
	% 		\hline
	% 		\hline
	%         Method & A-priori  & Ex-post   \\
	% 		\hline
	% 		 &  2.93~/~27.10 &  \bf{0.09~/~1.01}\\
	% 		\hline
	% 		\hline
	% 	\end{tabular}
	% 	\label{table_drpred}
	% \end{table}

	%Furthermore, Table~\ref{table_agentid} shows comparisons of baseline demand prediction based on the a-priori prediction (directly using the baseline demand forecast module) and the ex-post prediction (leverage the demand response module and net measurement) in more details. 

	\subsection{Real world demand response dataset}
	
\begin{figure}
		\centering
		\includegraphics[trim = 5mm 0mm 0mm 0mm, clip, width = .95\columnwidth]{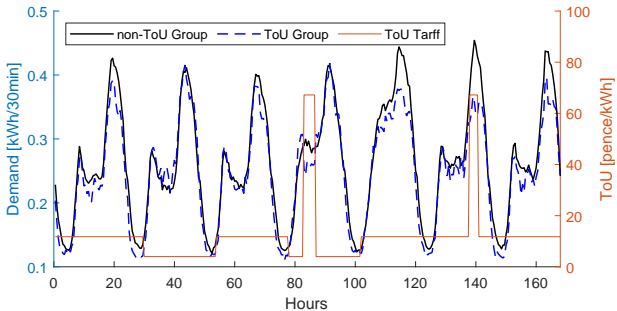}
		\caption{Example of average consumer demand from non-ToU and ToU groups.}
		\label{fig:london}
		\vspace{-3mm}
	\end{figure}

	We conduct identification experiments using real world demand response experiment data from the low carbon London project~\cite{london}, which included demand profile at 30-minute granularity during full year 2013 from 1,100 households receiving time-of-use (ToU) tariffs and 4,400 households receiving non-ToU flat rate (14.23 pence/kWh) for comparison, all located in the London area. The ToU tariff is set to 11.76 pence/kWh on default, 67.20 pence/kWh during high price periods, 3.99 pence/kWh during low price periods. High price periods were activated 69 times with a total duration of 394 hours, and low price periods were activated 93 times with a total duration of 830 hours. Customers are grouped into three categories based on income level: affluent, comfortable, and adversity, each category contains ToU and non-ToU customers. Fig.~\ref{fig:london} shows an example of the average consumer demand from the ToU group and the non-ToU group. 
	
	We design the agent optimization model as
	\begin{align}
	\textstyle  \min_{\{y_t\}} \sum_{t=1}^{T} \lambda_t y_t + \frac{0.5\alpha_1}{D_t-D\ud{min}} ([y_t]^+)^2 + \frac{0.5\alpha_2}{D_t-D\ud{min}} ([-y_t]^+)^2 %; M \geq 0
	\label{eq:london}
	\end{align}
	where $[x]^+ = \max\{0,x\}$ is the positive value function, $D_t$ is the baseline demand, and $D\ud{min}$ is the annual minimum demand that represents non-responsive demand, also called the ``ghost demand''. Hence, we assume the disutility coefficient is inverse proportional to the responsive baseline demand and has different values for demand increments ($\alpha_1$) and demand reduction ($\alpha_2$). We do not include the daily total response constraints in this design because  there is only at most one demand response event per day in the considered dataset.

	% We design the implicit layer for DR model considering that there are only three ToU price levels into the following formulation
	% \begin{align}\label{eq:london}
	% y_t = 
	%     \begin{cases}
	%     \alpha(D_t-D_{\mathrm{min}}) \quad & \text{if price is high} \\
	%     0 \quad & \text{if price is normal} \\
	%     \beta(D_t-D_{\mathrm{min}}) \quad & \text{if price is low} 
	%     \end{cases}
	% \end{align}
	% where $D_{\mathrm{min}}$ is the annual minimal demand of the group which can be considered as ``ghost demand'' that are not responsive to user behavior, $\alpha$ and $\beta$ are DR coefficients in which we assume the DR is proportional to the baseline demand $D_t$. This formulation provides an illustration of a demand-dependent DR agent model, and can be seen as an analytical solution to a quadratic DR optimization problem without constraints. We use this formulation directly due to the simple and infrequent DR periods in the dataset. %This DR model is connected to the MLP baseline demand prediction model to form the end-to-end framework to identify DR models using only ToU group demand data. 
	
	The input features to the baseline demand forecast MLP model are 6 dimensional: temperature, humidity, demand of the previous day and calendar information including hour/day/month. 
	%To improve the stability of the training, the model updates the DR coefficients $\alpha$ and $\beta$ at each training iteration and periodically updates the baseline forecast MLP parameters every 100 iterations. In addition, we update the previous day demand feature every 1000 iterations using the net demand minus the DR estimation.
	% \begin{table}[h]
	% 	\renewcommand{\arraystretch}{1.5}
	% 	\centering
	% 	\caption{Comparison between the E2E model identification result and benchmark model on London DR Dataset}
	% 	\begin{tabular}{lcccccc}
	% 		\hline
	% 		\hline
	%         \multirow{2}{4em}{Method} & \multicolumn{2}{c}{Affluent}  & \multicolumn{2}{c}{Comfortable} & \multicolumn{2}{c}{Adversity}\\
	%         & $\alpha$  & $\beta$   & $\alpha$   &  $\beta$ & $\alpha$ & $\beta$  \\
	% 		\hline
	%         E2E & -0.075 &  0.095 & -0.092 &  0.145 & -0.081 & 0.124\\
	%         Benchmark & -0.075 & 0.095 & -0.101 & 0.117 & -0.082 & 0.118 \\
	% 		\hline
	% 		\hline
	% 	\end{tabular}
	% 	\label{london_dr}
	% \end{table}
	\begin{table}[h]
		\renewcommand{\arraystretch}{1.5}
		\centering
		\caption{Comparison between the E2E model identification result and benchmark model on London DR Dataset}
		\begin{tabular}{lcccccc}
			\hline
			\hline
			\multirow{2}{4em}{Method} & \multicolumn{2}{c}{Affluent}  & \multicolumn{2}{c}{Comfortable} & \multicolumn{2}{c}{Adversity}\\
			& $\alpha_1$  & $\alpha_2$ &  $\alpha_1$  & $\alpha_2$ & $\alpha_1$  & $\alpha_2$  \\
			\hline
			E2E & 896 &  -42 & 730 &  -28 & 829 & -32\\
			Benchmark & 896 & -42 & 665 & -34 & 819 & -34 \\
			\hline
			\hline
		\end{tabular}
		\label{london_dr}
	\end{table}
	We compare our end-to-end model performance with a benchmark method to estimate $\alpha_1$ and $\alpha_2$ using control groups. The benchmark method follows a regression approach and features include non-ToU control group demand and temperature. The baseline prediction MAE of the benchmark method over the default tariff period is 0.009~kW. The MAPE is 5.8\%, $\alpha_1$ and $\alpha_2$ in \eqref{eq:london} is thus estimated using least square using the difference between the actual demand and the predicted baseline during the higher or low tariff period. 
	
	Table~\ref{london_dr} shows the estimation result of $\alpha_1$ and $\alpha_2$ as formulated in \eqref{eq:london} for three aggregated demand groups. The result shows our method generates very similar results compared to the benchmark method, which has access to the historical baseline demand information using a control group. Note that we do not assume any prior knowledge about the baseline demand or use demand profile from the non-ToU control group in the end-to-end method. Both methods provide very similar identification results and all are consistent with the study result from the London Project~\cite{london}. E2E training generates higher $\alpha_1$ in the Comfortable and Adversity group, possibly due to different error biases in the baseline prediction compared to the benchmark method; note that the benchmark method still has a baseline prediction MAPE of 5.8\%.
	% \subsection{Baseline estimation and demand prediction.}
	% \label{sec:result2}

	% As seen in Table~\ref{table_drpred}, both methods perform good in predicting the demand response, despite the ground-truth base load information. In addition, we found that the a-priori baseline estimation error (from the MLP load forecast module) is significant higher than the ex-post baseline estimation error (calculated by the net demand observation minus demand response forecast), which demonstrates the robustness of the implicit layer against model prediction errors. Fig~\ref{fig:prediction} provides an example of TM3 End-to-end training approach in A-priori baseline estimation, demand response forecast and ex-post baseline estimation in 5 consecutive test days. 
	
	\subsection{Ablation studies: additive Gaussian noise}
	% 1) model disturbance:  add synthetic  different noise  (Gaussain, random), test identification result; it’s good to analysis the true noise type from load prediction;
	% \textbf{Demand response identification with additive Gaussian noise:}
	As observed in Section~\ref{sec:result2}, the agent DR model identification module demonstrates robustness against baseline estimation errors, i.e., even though the a-priori baseline load estimation contains large errors, the demand response identification, and ex-post baseline estimation are accurate. In this section, we conduct investigative studies to understand why the OptNet layer is robust to the baseline estimation error, and how robust the model is against different levels of prediction errors. To exam the effect of prediction errors, we consider a setting where the incentive signal $\lambda_t$ and noisy demand response measurements $\hat{y}^{*}$ are available. In particular, we inject a zero-mean Gaussian noise with different standard deviations to the ground-truth demand response optima $y^{*}$. By changing the standard deviations of noise, we aim to simulate the effect of varying prediction errors, i.e., a larger standard deviation corresponds to a prediction model with higher errors.
	%It can be seen by comparing Fig~\ref{fig:prediction} (a) and Fig.~\ref{fig:data_vis}(b), although the a-priori baseline estimation error magnitude is significantly higher than the demand response magnitude, our proposed approach can achieve an accurate separation of user's demand response and baseline load.

	% For instance, Figure~\ref{fig:ablation_guassian} presents an example demand response groudtruth data with additive Gaussian noise $\mu = 0, \sigma = 1$.
	% \begin{figure}[htbp]%
	% 	\centering
	% 	\subfloat[]{
	% 		\includegraphics[trim = 0mm 00mm 0mm 0mm, clip, width = .95\columnwidth, height = .5\columnwidth]{figures/ablation1.eps}
	% 	}
	%   \caption{Visualization of additive white Gaussian noise to the ground-truth demand response data.}
	%   \label{fig:ablation_guassian}
	% \end{figure}

	\begin{table}[htbp]
		\renewcommand{\arraystretch}{1.5}
		\centering
		\caption{Demand response model identification error under different additive Gaussian noise.}
		\begin{tabular}{lcccccc}
			\hline
			\hline
			Parameter & $\sigma = 0$ & $\sigma = 0.5$ & $\sigma = 1$ & $\sigma = 2$ & $\sigma = 3$ & $\sigma = 5$\\
			\hline
			$\alpha$ (MAE) & 1.23e-5 &	0.062 	& 0.119	& 0.252	& 0.380 & 0.641 \\
			$\alpha$ (MAPE) & 0 &	0.26\% &	0.48\% &	1.04\% &	1.57\% & 2.64\%\\
			\hline
			$M$ (MAE) & 1.021e-4 &	0.159 &	0.258 &	0.636 &	0.955 & 1.591\\
			$M$ (MAPE) & 0 & 2.57\% & 4.17\% & 10.28\% & 15.43\% & 25.7\%\\
			\hline
			\hline
		\end{tabular}
		\label{table_ablation1}
	\end{table}
	Table~\ref{table_ablation1} shows the DR model identification results under different levels of Gaussian noise. Note that $\sigma = 0$ is the setting of no Gaussian noise. In practice, this corresponds to the scenario where the true baseline demands of users are available. As the noise level (i.e., $\sigma$) increases, the parameter identification error increases. But in general, the results demonstrate that the implicit optimization layer is robust to erroneous input data and can recover the underlying optimization problem parameters even under high-magnitude noise.
	
\begin{figure}[t]%
		\centering
		%	\subfloat[]{
		\includegraphics[trim = 0mm 5mm 0mm 5mm, clip, width = .9 \columnwidth]{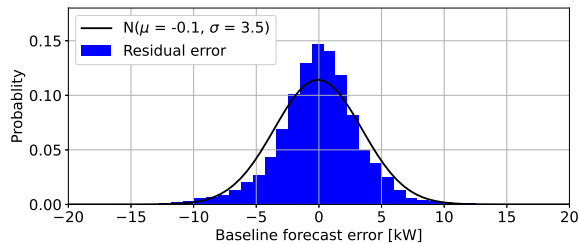}
		%	}
		\caption{Visualization of residual error of the baseline estimation module trained with historical baseline data. The residual error can be roughly fitted by a Gaussian distribution with mean $-0.1$ and standard deviation $3.5$.}
		\vspace{-3mm}
		\label{fig:ablation_guassian}
	\end{figure}

	In addition, we plot the residual error of the baseline demand forecast network trained with historical data in Section~\ref{sec:result2}. As shown in Fig~\ref{fig:ablation_guassian}, the residual error of the baseline demand forecast model roughly follows a normal distribution, where the mean is near zero (unbiased estimator), while the standard deviation equals $3.5$ (relatively high variance). The ablation result shows in Table~\ref{table_ablation1} that, the agent model can still achieve accurate parameter estimation, as long as the residual error of baseline estimation is near zero. Robustness of the implicit differential layer is an attractive property since it guarantees the applicability of the proposed approach under incomplete information (i.e., no true baseline), erroneous and noisy prediction, and measurement. An interesting future direction would be to formally study the robustness of the OptNet for DR model identification.
	
	% 2) how to decide what kind of model? what about model mismatch?
	%   —ground truth is linear model but use quadratic function to learn;
	%   ——the learned quadratic coefficient is small enough to ignore;
	%   or: constraint selection; (big constraint means no constraint)
	\subsection{Ablation studies: time-varying agent model}
	In all the previous experiments, the demand response and baseline models are trained with data collected, assuming the agent models are time-variant. However, in practice, users' preferences and flexibility for demand response may vary with respect to time. For instance, a user might be less willing to reduce the energy usage of his air conditioner on an extremely hot day than a mild day ($\alpha$ value is higher and $M$ value is smaller in the first case than the latter). This is challenging since the system operators don't know when/how the agent DR model is changed.
	
\begin{figure}[t]%
		\centering
		%	\subfloat[]{
		\includegraphics[trim = 0mm 5mm 0mm 5mm, clip, width = .9 \columnwidth]{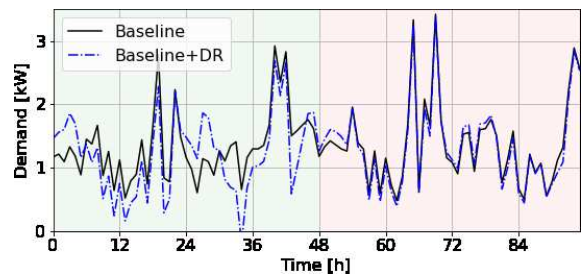}
		%	}
		\caption{Samples of baseline demand and net measurement with time-varying agent response model.}
		\label{fig:ablation2}
	\end{figure}

	We performed an experiment to mimic such effects whereby we generated data for training the demand response and baseline models from a mixture of two agent models. 
	As shown in Figure~\ref{fig:ablation2}, in the green area user model is $\alpha = 20, M=3$ (more responsive) and in the red area, user model is $\alpha = 50, M=0.5$ (less responsive). The identified result by our model is $\hat{\alpha} = 28.57, \hat{M} = 1.749$, which is close to the mean of the two models. It would be interesting to investigate different clustering or classification methods to first separate data from different user response modes and then run parameter identification for future works.
	
	% 4) [Optional] Performance of our model with or without baseline load (before DR enrollment) 

	\section{Conclusion}
	In this work, we developed an end-to-end deep learning model with to jointly identify demand response baseline and the agent decision making model. We proposed a gradient-descend based approach to solve the joint learning problem, and tested over both synthetic demand response and real-world demand response datasets. 
	Results show that our approach provides robust estimation performance against baseline prediction error and measurement noises. In our future work we plan to expand this learning framework with more sophisticated demand response agent models, and apply it over a dynamical identification setting such as rolling hour-ahead predictions and time-varying agent models. We are also interested in exploring the learning stability especially between the neural network and the implicit optimization model, and investigate more efficient training algorithm targeting learning models with more complex structures.

	\bibliographystyle{IEEEtran}
	\bibliography{main.bib}

	\appendix
	%  The KKT conditions of optimization problem~\eqref{eq:agent_model} can be written as, for $t = 1, ..., T$,
	% \begin{subequations}
	% \begin{align}
	%     \textstyle \lambda_t + \alpha_t y_t^{*} + \muo_t^{*} -\muu_t^{*} + \sum_{\tau=1}^t \nuo_{\tau}^{*} -  \sum_{\tau=1}^t \nuu_{\tau}^{*} & = 0 \\
	%     \muo_t^{*} (y_t^{*} - \overline{P})   = 0\,,\quad \muo_t^{*} &\geq 0 \\
	%     \muu_t^{*}  (\underline{P}-y_t^{*}) = 0\,,\quad \muu_t^{*} &\geq 0 \\
	%     \textstyle \nuo_t^{*} (\sum_{\tau=1}^t y_{\tau}^{*} - \overline{E})    = 0\,,\quad \nuo_t^{*} &\geq 0 \\
	%     \textstyle \nuu_t^{*} (\underline{E} - \sum_{\tau=1}^t y_{\tau}^{*})   = 0\,,\quad \nuu_t^{*} &\geq 0  
	% \end{align}
	% \end{subequations}
	% where \eqref{eq:kkt1} is the stationarity condition of the DR agent problem \eqref{eq:agent_model}, \eqref{eq:kkt2}--\eqref{eq:kkt5} are the complementary slackness conditions associated with constraints \eqref{eq:agent_contr1} and \eqref{eq:agent_contr2}. $y^{*}, \underline{\mu}^{*}, \overline{\mu}^{*}, \underline{\nu}^{*}, \overline{\nu}^{*}$ are the optimal primal and dual variables. 
	To derive the Jacobian terms of $y^{*}$ with respective to the agent DR model coefficients, we first re-write the DR agent optimization model~\eqref{eq:agent_model} in the compact matrix form,
	\begin{subequations}\label{eq:agent_model_matrix}
		\begin{align}
		\textstyle  \min_{y}\; g(D, y; \lambda, \theta) := & \lambda^\top (y+D) + \frac{\alpha}{2} y^\top y \,,\label{eq:agent_obj}\\
		\text{ s.t. } & I y \leq \overline{P}\,, \quad :\muo \label{eq:agent_power1}\\
		& -I y \leq -\underline{P}\,, \quad :  \muu\label{eq:agent_power2}\\
		& \Gamma y \leq \overline{E}\,, \quad :\nuo\label{eq:agent_energy1}\\
		& -\Gamma y \leq -\underline{E}\,, \quad :\nuu
		\label{eq:agent_energy2}
		\end{align}
	\end{subequations}
	
	The Lagrangian of~\eqref{eq:agent_model_matrix} is given by,
	\begin{align}
	L(y, \mu, \nu) &= \lambda^\top (y+D) + \frac{\alpha}{2} y^\top y + \overline{\mu}^\top(y- \mathbbm{1}\overline{P}) + \underline{\mu}^\top(\mathbbm{1}\underline{P}-y) \nonumber\\
	&+ \overline{\nu}^\top(\Gamma y-\mathbbm{1}\overline{E}) + \underline{\nu}^\top(\mathbbm{1}\underline{E}-\Gamma y)
	\end{align}
	where $\mathbbm{1}$ is an all-one vector and $\Gamma$ is a lower triangular matrix, with appropriate dimension.
	
	Then, we write out the KKT conditions for stationarity and complementary slackness conditions of the agent optimization problem in matrix form,
	\begin{subequations}
		\begin{align}
		\lambda + \alpha  y^{*} +  \overline{\mu}^{*} -\underline{\mu}^{*}+ \Gamma^\top \overline{\nu}^{*} - \Gamma^T \underline{\nu}^{*} &= 0 \\
		\Lambda(\overline{\mu}^{*})(y^* - \mathbbm{1}\overline{P}) &= 0\\
		\Lambda(\underline{\mu}^{*})(\mathbbm{1}\underline{P} - y^*) &= 0\\
		\Lambda(\overline{\nu}^{*})(\Gamma y^* - \mathbbm{1}\overline{E}) &= 0\\
		\Lambda(\underline{\nu}^{*})(\mathbbm{1}\underline{E} - \Gamma y^*) &= 0
		\end{align}
	\end{subequations}
	where $\Lambda(\cdot)$ creates a diagonal matrix from a vector. Taking the total derivatives of these conditions gives the following equations,
	\begin{subequations}\label{eq:total_derivative_kkt2}
		\begin{align}
		d \alpha y^{*} + \alpha dy + d \overline{\mu} - d \underline{\mu} + \Gamma^\top d \overline{\nu} - \Gamma^\top d \underline{\nu} &= 0 \\
		\Lambda(y^{*}-\mathbbm{1}\overline{P}) d\overline{\mu} + \Lambda(\overline{\mu}^{*})(dy - \mathbbm{1}d\overline{P}) &= 0\\
		\Lambda(\mathbbm{1}\underline{P}-y^{*}) d\underline{\mu} + \Lambda(\underline{\mu}^{*})(\mathbbm{1}d\underline{P}-dy) &= 0\\
		\Lambda(\Gamma y^{*}-\mathbbm{1}\overline{E}) d\overline{\nu} + \Lambda(\overline{\nu}^{*})(\Gamma dy - \mathbbm{1}d\overline{E}) &= 0\\
		\Lambda(\mathbbm{1}\underline{E}-\Gamma y^{*}) d\underline{\nu} + \Lambda(\underline{\nu}^{*})(\mathbbm{1}d\underline{E}-\Gamma dy) &= 0
		\end{align}
	\end{subequations}
	We re-write Eq~\eqref{eq:total_derivative_kkt2} in the compact matrix form and obtain \eqref{eq:total_derivative_kkt_compact} in the main paper.
\end{document}